\documentclass[conference, 10pt]{IEEEtran}
  \usepackage[absolute]{textpos}
\newcommand{\copyrightstatement}{
    \begin{textblock}{15}(0.5,0.3)    
         \noindent
         \centering
         \textblockcolour{white}
         \footnotesize
         \copyright 2018 IEEE. Personal use of this material is permitted. Permission from IEEE must be obtained for all other uses, in any current or future media, including reprinting/republishing this material for advertising or promotional purposes, creating new collective works, for resale or redistribution to servers or lists, or reuse of any copyrighted component of this work in other works
    \end{textblock}
}

\usepackage{cite}
\usepackage{url}

\ifCLASSINFOpdf
\else
 \usepackage[dvips]{graphicx}
  \DeclareGraphicsExtensions{.eps}
\fi

\usepackage{psfrag}

\usepackage[cmex10]{amsmath}

\usepackage{graphicx}
\usepackage{color}
\usepackage{float}

\begin{document}

\copyrightstatement

%
\title{Improving HEVC Encoding of Rendered Video Data Using True Motion Information}

\author{\IEEEauthorblockN{Christian Herglotz$^1$, David M\"uller$^2$, Andreas Weinlich$^1$, Frank Bauer$^3$, Michael Ortner$^1$, \\ Marc Stamminger$^3$, Andr\'e Kaup$^1$}
\IEEEauthorblockA{$^1$Chair of Multimedia Communications and Signal Processing, 
Friedrich-Alexander University Erlangen-N\"urnberg (FAU)\\
\{christian.herglotz, andreas.weinlich, michael.ortner, andre.kaup\}@fau.de \\ 
$^2$ Chair of Applied Informatics V, University of Bayreuth, david.mueller@uni-bayreuth.de \\
$^3$Chair of Computer Science 9, Friedrich-Alexander University Erlangen-N\"urnberg (FAU)\\
bauer@cs.fau.de, marc.stamminger@fau.de
}}

%
%
%




\maketitle

\begin{abstract}
This paper shows that motion vectors representing the true motion of an object in a scene can be exploited to improve the encoding process of computer generated video sequences. Therefore, a set of sequences is presented for which the true motion vectors of the corresponding objects were generated on a per-pixel basis during the rendering process. 
In addition to conventional motion estimation methods, it is proposed to exploit the computer generated motion vectors to enhance the rate-distortion performance. To this end, a motion vector mapping method including disocclusion handling is presented. It is shown that mean rate savings of $3.78\%$ can be achieved.  
\end{abstract}


\IEEEpeerreviewmaketitle

\section{Introduction}
\label{sec:intro}
With to the establishment of smartphones in the past decade, video streaming and online gaming applications have become popular all over the world. Nowadays, more than half of the global internet traffic is constituted of video data \cite{cisco17}. In this respect, the latest video coding standard high-efficiency video coding (HEVC) \cite{Sullivan12} is one of the most widely used codecs. 

A major challenge in video compression is the encoding process. The goal of this process is to compress a given input sequence with a small bitrate such that transmission channel capacities are met. Counteractingly, the distortion should be as small as possible. Therefore, in the so-called rate-distortion optimization (RDO) \cite{Sullivan98} during encoding, a large amount of coding modes is tested to find the one that leads to the lowest distortion at a given bitrate. 
In this process, one of the computationally most complex tasks is the search for the motion vector (MV) in interframe prediction. In a frame of a high-definition (HD) sequence, the search space includes more than $33$ million MVs pointing to a single reference frame 
(including quarter pixel MVs in HEVC). 

In order to reduce the computational complexity while still obtaining reasonable rate-distortion (RD) performance, the search space is pruned using sophisticated search patterns. For the search pattern in the HM software \cite{HM}, which is used as a reference, the motion estimation stage requires $66\%$ of the total encoding time ($17\%$ for the integer pel and $49\%$ for the fractional pel motion estimation \cite{Vanne12}). In the literature, further search methods can be found. 
For HEVC, search patterns were refined by testing adaptive search ranges \cite{Kibeya16} or allowing asymmetric search patterns \cite{Nguyen14}. Another search simplification method relies on additional video data. For example, Lee et al. exploit depth information  \cite{tsz-kwan17} to adapt the search range. All these search methods mainly result in savings in terms of encoding time, however the bitrate usually increases by several percent. 

\begin{figure}
\centering
\psfrag{R}[c][c]{Renderer}
\psfrag{V}[c][c]{Vectors}
\psfrag{S}[c][c]{Sequence}
\psfrag{M}[c][c]{Motion}
\psfrag{D}[c][c]{Depth}
\psfrag{W}[c][c]{Map}
\psfrag{E}[c][c]{Encoder}
\psfrag{B}[c][c]{Bit stream}
\includegraphics[width=.3\textwidth]{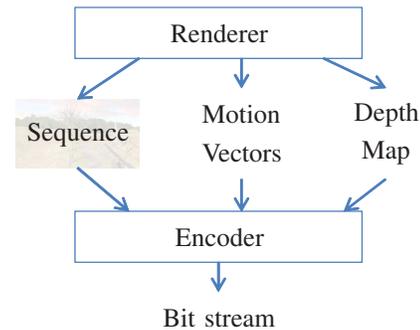}
\vspace{-.5cm}
\caption{Block diagram of rendering and encoding pipeline.}
\label{fig:setup}
\end{figure}

In this paper, a new motion search method is presented that makes use of precomputed motion vectors and depth maps. That means that the encoder receives additional information as shown in Fig.\ \ref{fig:setup}. Next to the input sequence, a motion vector field (MVF) and a depth map (DM) are provided for each frame. In this paper, the case of computer generated sequences is considered because for such sequences, the MVF and the DM can be calculated during the rendering process. To this end, the true motion of an object is projected to the image plane and saved on a per-pixel basis. Here, a special focus is put on online gaming scenarios in which short latency is important. 
As a consequence, the case of lowdelay\_P coding is considered \cite{Bossen13} in which 
 only the preceding frame is used for motion compensation. 
Thus, the feedback time is short and an exhaustive motion search for multiple reference frames is skipped. 

The modified HEVC encoder is presented in Section \ref{sec:enc}. Afterwards, Section \ref{sec:seq} presents various test sequences with annotated motion information that was created using Blender \cite{Blender}. Section \ref{sec:eval} evaluates the performance of the proposed encoder in terms of rate, distortion, and encoding time. Finally, Section \ref{sec:concl} concludes this paper.



\section{Encoder Modifications}
\label{sec:enc}

To exploit the additional information given by the motion vector field and the depth map, the HM encoder version 12.0 \cite{HM} is modified. 
The method extends the conventional motion estimation on prediction unit (PU) level in HM, which performs two searches: In the first search, an eight-point diamond search is performed on an integer pel basis. 
Second, a fractional pel refinement is processed to achieve quarter-pel accuracy. 
The proposed method is an additional search on top of these two conventional methods which tests motion vectors that were not tested before.

\begin{figure}
\centering
\psfrag{D}[l][l]{Disocclusion}
\psfrag{H}[l][l]{Handling}
\psfrag{A}[l][l]{\small{Pixel Grid}}
\psfrag{B}[l][l]{\small{Partition Borders}}
\psfrag{C}[l][l]{\small{PU Borders}}
\psfrag{E}[l][l]{\small{Motion Vectors}}
\psfrag{M}[c][c]{Motion Vector}
\psfrag{V}[c][c]{Mapping}
\psfrag{S}[c][c]{MV Testing}
\includegraphics[width=.48\textwidth]{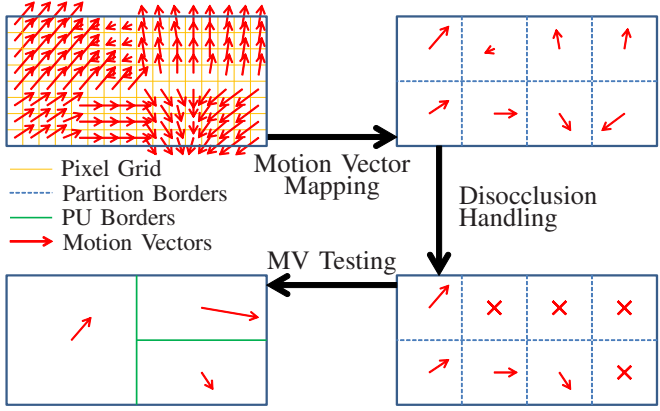}
\vspace{-.3cm}
\caption{Proposed Motion Vector Search Method with the three stages motion vector mapping, disocclusion handling, and motion vector testing. The red 'x'es denote invalid motion vectors as detected by disocclusion handling. }
\label{fig:block_diag_vectors}
\end{figure}

 The proposed search method consists of the three major stages motion vector mapping, disocclusion handling, and motion vector testing as depicted in Fig.\ \ref{fig:block_diag_vectors}. The stages are explained in the following sections.  

\subsection{Block-Wise Motion Vector Mapping}
\label{sec:motRetrieve}
As a first step, 
a mapping method from pixel-wise to block-wise motion vectors is introduced to prune the search space. 
The complete frame is divided in blocks of $4\times 4$ pixels which corresponds to the minimum partition size used in HEVC. For each block, the following steps are taken: 
\begin{enumerate}
\item The x-median and the y-median of all motion vectors are determined.
\item Both vectors containing one of the median values are used as candidates.
\item For both candidates, the sum of squared differences to all $16$ motion vectors in the block is calculated. 
\item The motion vector with the smaller squared error is chosen as the representative motion vector. 
\item If the motion vector points outside of the frame, it is considered unusable and marked invalid. 
\end{enumerate}

The median value of the motion vectors is taken because it represents the true motion for at least one pixel. Taking, e.g., the mean of the motion vectors would result in an artificial motion vector that can be invalid for all the pixels. 

\subsection{Disocclusion Handling}

\begin{figure}
\centering
\psfrag{C}[b][t]{Camera}
\psfrag{F}[c][c]{\underline{Previous Frame}}
\psfrag{G}[c][c]{\underline{Current Frame}}
\psfrag{z}[l][l]{\color[rgb]{1,0,0}$\boldsymbol{m}(x,y)$}
\psfrag{v}[c][c]{\color[rgb]{1,0,0}$\boldsymbol{p}$}
\psfrag{w}[c][c]{\color[rgb]{1,0,0}$\boldsymbol{p}_\mathrm{prev} $}
\includegraphics[width=.4\textwidth]{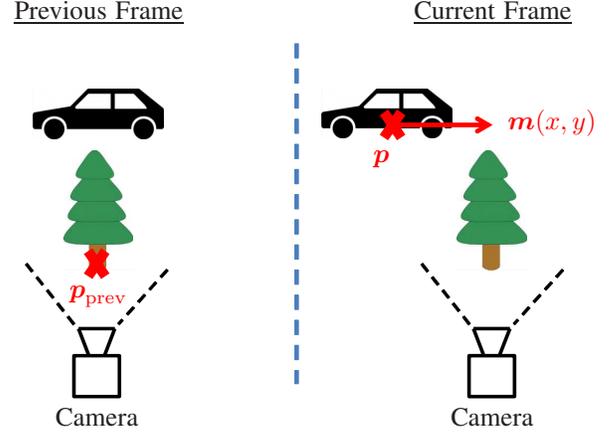}
\vspace{-.8cm}
\caption{Visualization of a disocclusion calculation. In the previous frame, the car is occluded by the tree. In the current frame, the car is visible. Using the motion $\boldsymbol{m}(x,y)$ of the car, one can check whether the car was visible in the preceding frame. }
\label{fig:disoccl}
\end{figure}

A disocclusion occurs if a visible object in the current frame is occluded in the previous frame. 
Such a case is visualized in Fig. \ref{fig:disoccl}. 
In the previous frame, the car is occluded, 
in the current frame, it is visible. 
Although a motion vector $\boldsymbol{m}(x,y)$ can be calculated for the car (as shown in the current frame), the vector cannot be used for prediction because the visual information taken from the previous frame would refer to a different object (the tree at position $\boldsymbol{p}_\mathrm{prev}$). 

Mathematically, the disocclusion can be detected as follows. Consider we check the pixel at position $x$ and $y$. Then, we can define the position of this pixel in the current frame and the position of the corresponding pixel in the previous frame as
\begin{equation}
\boldsymbol{p} = \left( \begin{array}{c} x\\y\\z(x,y)\end{array} \right), \ \boldsymbol{p}_\mathrm{prev} = \left( \begin{array}{c} x_\mathrm{prev}\\y_\mathrm{prev}\\z_\mathrm{prev}(x_\mathrm{prev},y_\mathrm{prev})\end{array} \right), 
 \label{eq:p_curr}
\end{equation}
where the third coordinates (i.e. the depths $z(x,y)$ and $z_\mathrm{prev}$) denote the distance of the visible object to the camera as indicated by the input DMs. $x_\mathrm{prev}$, $y_\mathrm{prev}$, and $z_\mathrm{prev}$ describe the corresponding position in the previous frame. $x_\mathrm{prev}$, $y_\mathrm{prev}$ and the test-depth $z_\mathrm{test}$ can be calculated by adding the 3D motion $\boldsymbol{m}(x,y)$ to the current position $\boldsymbol{p}$. \begin{equation}
\left( \begin{array}{c} x_\mathrm{prev}\\ y_\mathrm{prev}\\ z_\mathrm{test}\end{array} \right)=\boldsymbol{p} + \boldsymbol{m}(x,y).  
\label{eq:prev_pos_calc}
\end{equation}
Then, using the position ($x_\mathrm{prev},y_\mathrm{prev}$), the depth of the displayed object in the previous frame $z_\mathrm{prev}$ can be compared to the depth of the current object in the previous frame $z_\mathrm{test}$ as
\begin{equation}
	 z_\mathrm{test} > z_\mathrm{prev}( x_\mathrm{prev}, y_\mathrm{prev})+T. 
	 \label{eq:disoccl}
\end{equation}
A disocclusion occurs if this inequation holds because the current object is occluded by, i.e. located behind another object in the previous frame.

The threshold $T$ is used to prevent depth-fighting \cite{Vasilakis13} that can occur when no disocclusion happens. Because of numerical inaccuracies, the signaled position $z_\mathrm{prev}$ can differ slightly from the calculated position $z_\mathrm{test}$ although it belongs to the same object. Hence, to avoid falsely detected disocclusions in the case that $z_\mathrm{test}$ is just slightly greater than $z_\mathrm{prev}$, the threshold $T$ is introduced. In this work, $T=0.004$ is chosen because it proved to yield good results.
 
The disocclusion check is performed on a $4\times 4$-block basis. Therefore, each of the $16$ pixels are tested. If more than $8$ pixels are disoccluded, more than half of the block includes new visual information. It is assumed that as a consequence, traditional motion search is more efficient than the proposed search method and the motion vector derived by motion vector mapping is marked invalid.

\subsection{Motion Vector Testing}
In HM, the motion estimation is performed on prediction unit (PU) level which consists of two to maximum $256$ blocks of size $4\times 4$. 
From each of these blocks, the valid motion vectors are stored in a list in which double entries are removed. Additionally, the two motion vectors from motion vector prediction (MVP) \cite{Sullivan12} are added to the list. Then, all motion vectors in the list 
are checked for distortion. In contrast to the original motion search in which the best MVP is determined after finding the best motion vector, in the proposed search the best MVP is directly used to calculate rate costs. 
Then, the best vector is chosen from both the conventional and the proposed search and is coded into the bit stream. 

\section{Computer Generated Sequences}
\label{sec:seq}

For evaluation, eleven sequences were rendered with Blender. 
For each sequence, depth and motion information was determined during rendering by projecting the physical 3D motion to the 2D image plane. In x- and y-dimension, the motion is calculated in quarter pel accuracy, the z-motion is given in the depth domain. 
The motion vectors are calculated with respect to the last frame. The resulting information is transmitted to the encoder. 

\begin{figure}
\centering
\psfrag{A}[c][c]{ArrowsCIF}
\psfrag{B}[c][c]{HexaCIF}
\psfrag{C}[c][c]{PilotJumpCIF}
\psfrag{D}[c][c]{CountrySoldier}
\psfrag{E}[c][c]{Carshow}
\psfrag{F}[c][c]{CarChecker}
\includegraphics[width=.4\textwidth]{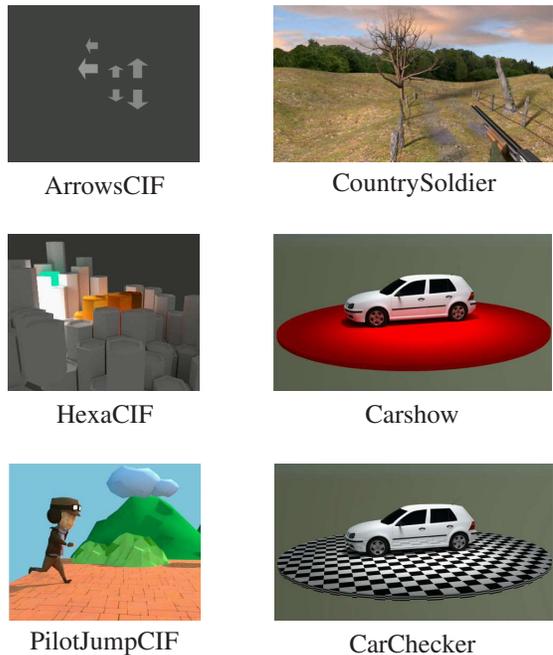}
\vspace{-.4cm}
\caption{Example frames of six different test sequences generated with Blender. }
\label{fig:seq_example}
\end{figure}

Visual examples of the sequences are shown in Fig. \ref{fig:seq_example}. The first type of sequence considers multiple moving arrows in a 2D world (ArrowsCIF and ArrowsHD). The arrows move in different directions at variable speeds. 

The second type of sequence shows various hexagonal tubes moving upwards and downwards in 3D. Some of them radiate light, some only reflect incoming light. 
Two sequences are given with a fixed camera position (HexaCIF and HexaHD). Furthermore, two more sequences are tested with a moving camera at HD resolution. To show differences in render engines, one of the sequences is rendered with Blender's standard render engine (HexaStand) and one is rendered with the ray-tracer Cycles \cite{Blender} (HexaCyc). 

The third type of sequences shows a 3D-Jump-and-Run scenario typical for a computer game. A pilot is running through a 3D-world (PilotJumpCif and PilotJumpHD). 
 The fourth type considers a first-person shooter game in a realistic countryside (CountrySolider). The last type is a virtual car-showroom in which two cars are presented on a platform (Carshow and CarChecker). The latter sequence uses a checkerboard texture on the platform which is reflected of the lacquer of the car. The visual motion of these reflections is different from the true motion of the car. 

In total, $11$ sequences were tested. The main properties are summarized in Table \ref{tab:sequences}. The sequences and their motion information can be downloaded at \cite{motionSeqs} and are free for personal use. 
\begin{table}[t]
\renewcommand{\arraystretch}{1.3}
\caption{Evaluation sequences and major properties. }
\label{tab:sequences}
\begin{center}
\vspace{-0.6cm}
\begin{tabular}{l|c|c|c}
\hline
Name & \# frames & Resolution & Renderer\\
 \hline
ArrowsCIF & $15$ & CIF &  standard \\
ArrowsHD & $15$ & HD & standard \\
HexaCIF & $60$ & CIF &  Cycles \\
HexaHD & $60$ & HD &  Cycles \\
HexaStand & $54$ & HD &  standard \\
HexaCyc & $54$ & HD &  Cycles \\
PilotJumpCIF & $30$ & CIF &  standard \\
PilotJumpHD & $30$ & HD &  standard \\
CountrySoldier & $200$ & HD & standard \\
Carshow & $250$ & HD &  Cycles \\
CarChecker & $250$ & HD & Cycles \\
\hline
\end{tabular}
\end{center}
\vspace{-0.3cm}
\end{table}

\section{Evaluation}
\label{sec:eval}
In this section, the rate-distortion performance of the proposed motion search method is tested for all sequences introduced in Section \ref{sec:seq}. The reference is the conventional case implemented in the standard HM encoder with lowdelay{\_}P configuration. Besides, the four QPs $22$, $27$, $32$, and $37$ are tested. Only the motion search algorithm is changed, the search for other modes like intra, merge, or skip mode were not modified. Furthermore, the common inter search range of $64$ pixels is kept which means that when higher motion vector changes occur in a sequence, the proposed method has an advantage because it has no restrictions on the search range. 
Furthermore, all standard encoder speedup methods (e.g., early skip decision) are enabled for all coding orders by default. 

We evaluate the distortion in terms of YUV-PSNR  \cite{Ohm12}.
The rate is evaluated in terms of bit stream file size in kilobytes (kB). As a third evaluation criterion, the encoding time is studied. Therefore, the output of the C++-\texttt{clock()} function is used. Encoding is performed single-thread on an Intel Core i7-3770 at $3.4$ GHz with $16$ GB RAM. The operating system is Windows 7, $64$ bit. 
For the evaluation of the encoding time, the reading of the motion information from hard disk is not taken into account because in a practical application, the data could be directly transmitted from the renderer to the encoder. 

We use the Bj{\o}ntegaard-Delta rates \cite{Bjonte01} and the relative encoding time differences for analysis.   The values are listed in Table \ref{tab:results_best_orders}. 
\begin{table}[t]
\renewcommand{\arraystretch}{1.3}
\caption{RD-performance and encoding time savings $\Delta T$ of the proposed search method in comparison to the conventional method.  }
\label{tab:results_best_orders}
\begin{center}
\vspace{-0.4cm}
\begin{tabular}{l||r|r}
\hline
 Sequence & BD-Rate & $\Delta T$\\
 \hline
ArrowsCIF & \color[rgb]{0,.5,0}$ -3.22\%$ & $ 1.44\%$ \\ 
ArrowsHD & \color[rgb]{0,.5,0}$ -23.4\%$ & $ 3.94\%$ \\ 
HexaCIF  & \color[rgb]{0,.5,0}$ -0.446\%$ & $ 9.48\%$ \\ 
HexaHD & \color[rgb]{0,.5,0}$ -1.37\%$ & $ 8.81\%$ \\ 
HexaStand  & \color[rgb]{0,.5,0}$ -3.68\%$ & $ 18.5\%$ \\ 
HexaCyc & \color[rgb]{0,.5,0}$ -2.16\%$ & $ 16.9\%$ \\ 
PilotJumpCIF & \color[rgb]{0,.5,0}$ -0.97\%$ & $ 7.42\%$ \\ 
PilotJumpHD & \color[rgb]{0,.5,0}$ -4.63\%$ & $ 1.46\%$ \\ 
CountrySoldier & \color[rgb]{0,.5,0}$ -0.634\%$ & $ 6.42\%$ \\ 
Carshow & \color[rgb]{0,.5,0}$ -0.615\%$ & $ 14.6\%$ \\ 
CarChecker  & \color[rgb]{0,.5,0}$ -0.462\%$ & $ 10.5\%$ \\ 
\hline
Average &  \color[rgb]{0,.5,0}$ -3.78\%$ & $ 9.04\%$ \\ 
\hline
\end{tabular}
\end{center}
\vspace{-0.6cm}
\end{table}
The table shows that for all sequences, significant rate savings can be achieved ($-0.446\%$ to $-4.63\%$). One sequence shows extreme rate savings (ArrowsHD with more than $20\%$) which is caused by true motion vectors being larger than the search range. As a result, the conventional search does not find a good prediction such that more residual coefficients must be coded. On average, $3.78\%$ of rate could be saved. 

In terms of complexity, we can see that the encoding time increases by $9.04\%$ on average (see Table \ref{tab:results_best_orders}). 
The increase is highest for sequences with heterogeneous motion (HexaStand, HexaCyc, Carshow, and CarChecker). The reason is that especially for large blocks, more motion vectors must be tested. Sequences with homogeneous motion (ArrowsCIF and ArrowsHD) only have small encoding time increases because even for large blocks, only a couple of different true motion vectors exist.

\section{Conclusions}
\label{sec:concl}
This paper showed that motion vectors representing true motion, which can be calculated during the rendering process of computer generated sequences, can be exploited to improve the compression performance. To this end, a motion vector mapping from pixel to block based motion vectors with subsequent motion vector testing is proposed. Performing this search, average rate savings of $3.78\%$ can be generated. 
Note that generally, the concept is applicable to any modern video codec that makes use of motion compensation. 

Future work could optimize the motion vector mapping method or prune the amount of tested motion vectors. First tests showed that when the traditional search is replaced with the proposed search, the encoding time can be decreased by $13.9 \%$, however at an increased BD-rate of $12.2\%$. Another approach is to concentrate on the handling of reflections and illumination changes.  Furthermore, by providing motion vectors to multiple sequences, the lowdelay\_B and the randomaccess configurations could also be considered.  
Finally, the use of precalculated motion vectors allows for affine and homographic instead of translational motion estimation with almost no additional encoding complexity. 


 
\bibliographystyle{IEEEtran}
\bibliography{IEEEabrv,literatureNeu}

\end{document}